\documentclass[showpacs,preprintnumbers,amsmath,amssymb,floatfix]{revtex4}
\usepackage{epsf}
 \newcommand{\insertplot}[5]{\begin{figure}
 \hfill\hbox to 0.05in{\vbox to #5in{\vfill
 \inputplot{#1}{#4}{#5}}\hfill}
 \hfill\vspace{-.1in}
 \caption{#2}\label{#3}
 \end{figure}}
 \newcommand{\inputplot}[3]{
 \special{ps: plotfile #1}
}

\usepackage[german, english]{babel}
\usepackage{ifthen}
\usepackage{epsfig}
\newcounter{fig}   \newcommand{\lbfig}[1]{\refstepcounter{fig}
\label{#1} }

\newcommand{\vphi}{\varphi}

\begin{document}
\title{Gravitating Sphaleron-Antisphaleron Systems}
\author{
{\bf Rustam Ibadov}
}
\affiliation{Department of Theoretical Physics and Computer Science,\\
Samarkand State University, Samarkand, Uzbekistan}
\author{
{\bf Burkhard Kleihaus, Jutta Kunz and Michael Lei\ss ner}
}
\affiliation{
{Institut f\"ur Physik, Universit\"at Oldenburg, 
D-26111 Oldenburg, Germany}
}
\date{\today}

\begin{abstract}
We present new classical solutions of Einstein-Yang-Mills-Higgs theory,
representing gravitating sphaleron-antisphaleron pair, chain and vortex ring 
solutions. In these static axially symmetric solutions, 
the Higgs field vanishes on isolated points on the symmetry axis, 
or on rings centered around the symmetry axis.
We compare these solutions to gravitating monopole-antimonopole
systems, associating monopole-antimonopole pairs with sphalerons.
\end{abstract}

\maketitle

\section{Introduction}

The non-trivial topology of the configuration space of the 
bosonic sector of the SU(2)$\times$U(1) electroweak theory
gives rise to a plethora of unstable classical solutions. 
Besides the Klinkhamer-Manton sphaleron \cite{km,kkb},
Weinberg-Salam theory also allows for multisphalerons,
which possess either axial or platonic symmetries \cite{multi,Kari}.
Moreover sphaleron-antisphaleron pairs are present \cite{klink}, 
and, as shown recently, also sphaleron-antisphaleron
chains and vortex rings \cite{kkl}.

When gravity is coupled to this Yang-Mills-Higgs (YMH) theory,
the flat space sphaleron changes smoothly, and a branch
of gravitating sphalerons arises \cite{greene,vg,yves}.
This branch bifurcates at a maximal value of the gravitational
coupling constant with a second branch, higher in energy.
In the limit of vanishing coupling constant,
this second branch ends at the lowest 
Bartnik-McKinnon (BM) solution \cite{bm}
of Einstein-Yang-Mills (EYM) theory.

Here we consider the effect of gravity on the
axially symmetric multisphalerons, and the
sphaleron-antisphaleron pairs, chains and vortex rings.
We characterize these solutions by two integers, $m$ and $n$.
The Klinkhamer-Manton sphaleron has $m=n=1$,
while the multisphalerons, representing
superpositions of $n$ sphalerons, have $m=1,n>1$.
Sphaleron-antisphaleron pairs are obtained for $m=2,n=1,2$,
and chains for $m>2,n=1,2$, while vortex rings arise for $m>1,n>2$.

We find, that all these unstable gravitating solutions
show the same general coupling constant dependence,
as observed for the single gravitating sphaleron.
Two branches of solutions arise, a lower branch connected
to the flat space solution and an upper branch connected to
an EYM solution.
For $m = 1$ these EYM solutions
correspond to the spherically symmetric BM solution ($n=1$)
or their axially symmetric generalizations ($n>1$) \cite{bm,kk},
whereas for $m \ge 2$ ($n \ge 4$)
these limiting EYM solutions are of a different type \cite{ikks}.
At the same time additional branches of solutions arise,
which connect to the generalized BM solutions \cite{kk}.

All these solutions thus show a gravity dependence,
similar to the monopole-antimonopole pair, 
chain and vortex ring solutions
\cite{map,kks-g}, encountered in the Georgi-Glashow model
coupled to gravity,
where the Higgs field is not in the fundamental representation
of SU(2), but instead in the adjoint representation.
We therefore here address the analogy of both sets of solutions.
In particular, we find major agreement concerning their general pattern,
when we compare sphalerons and sphaleron-antisphaleron systems,
characterized by the integers $m$ and $n$,
with monopole-antimonopole systems,
characterized by the integers $2m$ and $n$.

In section II we briefly review the action of SU(2) EYMH theory.
We present the static axially symmetric Ans\"atze 
and the boundary conditions for the solutions in section III.
In section IV we then present our numerical results
for sphaleron-antisphaleron pairs, chains and vortex rings,
and discuss the physical properties of these solutions.
We give our conclusions in section V.

\section{Einstein-Yang-Mills-Higgs Theory}

We consider SU(2) Einstein-Yang-Mills-Higgs theory with action
\begin{equation}
S =  \int \left\{ \frac{R}{16\pi G}
-\frac{1}{2} {\rm Tr}
\,\left( F_{\mu\nu}F^{\mu\nu} \right)
- (D_\mu \Phi)^{\dagger} (D^\mu \Phi)
- \lambda (\Phi^{\dagger} \Phi - \frac{v^2}{2} )^2
\right\}
\sqrt{- g} d^4 x
\ , \label{action} 
\end{equation}
with curvature scalar $R$,
su(2) field strength tensor
\begin{equation}
F_{\mu\nu}=\partial_\mu V_\nu-\partial_\nu V_\mu
            + i e [V_\mu , V_\nu ]
\ , \end{equation}
su(2) gauge potential $V_\mu = V_\mu^a \tau_a/2$,
and covariant derivative of the Higgs $\Phi$
in the fundamental representation
\begin{equation}
D_{\mu} \Phi = \Bigl(\partial_{\mu}
             +i e  V_{\mu}  \Bigr)\Phi
\ , \end{equation}
where $G$ and $e$ denote the gravitational and gauge coupling constants,
respectively,
$\lambda$ denotes the strength of the Higgs self-interaction and
$v$ the vacuum expectation value of the Higgs field.

The action (\ref{action}) is invariant under local $SU(2)$
gauge transformations $U$,
\begin{eqnarray}
V_\mu &\longrightarrow & U V_\mu U^\dagger
+ \frac{i}{e} \partial_\mu U  U^\dagger \ ,
\nonumber\\
\Phi  &\longrightarrow & U \Phi\ .
\nonumber
\end{eqnarray}
The gauge symmetry is spontaneously broken 
due to the non-vanishing vacuum expectation
value of the Higgs field
\begin{equation}
    \langle \Phi \rangle = \frac{v}{\sqrt2}
    \left( \begin{array}{c} 0\\1  \end{array} \right)   
\ , \end{equation}
leading to the vector and scalar boson masses
\begin{equation}
    M_W = \frac{1}{2} e v \ , \ \ \ \ 
    M_H = v \sqrt{2 \lambda} \ . 
\end{equation}

In the absence of gravity, the Lagrangian reduces to the
bosonic sector of Weinberg-Salam theory with vanishing Weinberg angle.
Setting the Weinberg angle to zero
is a good approximation for sphalerons and multisphalerons \cite{multi}.
We here consider gravitating sphaleron, sphaleron-antisphaleron chain
and vortex ring solutions in the limit of vanishing Weinberg angle.

In the standard model fermion number is not conserved \cite{thooft}.
Reexpressing the anomaly term in terms of the Chern-Simons current
\begin{equation}
 K^\mu=\frac{e^2}{16\pi^2}\varepsilon^{\mu\nu\rho\sigma} {\rm Tr}(
 F_{\nu\rho}V_\sigma
 + \frac{2}{3} i e V_\nu V_\rho V_\sigma )
\ , \end{equation}
yields for the fermion charge of a sphaleron solution
(in a suitable gauge) \cite{km}
\begin{equation}
 Q_{\rm F} = \int d^3r K^0 \ .
\label{Q}
\end{equation}

\section{Ansatz and boundary conditions}

To obtain gravitating static axially symmetric solutions,
we employ isotropic coordinates \cite{kk}.
In terms of the spherical coordinates $r$, $\theta$ and $\vphi$
the isotropic metric reads
\begin{equation}
ds^2=
  - f dt^2 +  \frac{h}{f} d r^2 + \frac{h r^2}{f} d \theta^2
           +  \frac{l r^2 \sin^2 \theta}{f} d\vphi^2
\ , \label{metric2} \end{equation}
where the metric functions
$f$, $h$ and $l$ are functions of
the coordinates $r$ and $\theta$, only.
The $z$-axis ($\theta=0, \pi$) represents the symmetry axis.
Regularity on this axis requires
\begin{equation}
h|_{\theta=0, \pi}=l|_{\theta=0, \pi}
\ . \label{lm} \end{equation}

We take a purely magnetic gauge field, $V_0=0$,
and parametrize the gauge potential and the Higgs field by the Ansatz
\cite{kks,kkl}
\begin{equation}
V_i dx^i = \left(\frac{H_1}{r} dr + (1-H_2) d\theta\right)
           \frac{\tau^{(n)}_\vphi}{2e}
          -n\sin\theta\left(H_3 \frac{\tau^{(n,m)}_r}{2e} 
	  + (1-H_4)\frac{\tau^{(n,m)}_\theta}{2e}\right) d\vphi
	   \ , \ \ \ V_0=0 \ ,
\label{a_axsym}
\end{equation}	  
and
\begin{equation}
\Phi = i( \Phi_1 \tau^{(n,m)}_r + \Phi_2 \tau^{(n,m)}_\theta )\frac{v}{\sqrt2}
    \left( \begin{array}{c} 0\\1  \end{array} \right) \ .
\end{equation}
where
\begin{eqnarray}	  
\tau^{(n,m)}_r & = & \sin m\theta (\cos n\vphi \tau_x + \sin n\vphi \tau_y) 
           + \cos m\theta \tau_z \ , \ \ 
\nonumber \\	   
\tau^{(n,m)}_\theta & = & \cos m\theta (\cos n\vphi \tau_x + \sin n\vphi \tau_y) 
           - \sin m\theta \tau_z \ , \ \ 
\nonumber \\	   
\tau^{(n)}_\vphi & = & (-\sin n\vphi \tau_x + \cos n\vphi \tau_y) 
\ , \ \ \nonumber 
\end{eqnarray}	  
$n$ and $m$ are integers, 
and $\tau_x$, $\tau_y$ and $\tau_z$ denote the Pauli matrices.

The two integers $n$ and $m$ determine the fermion number of the
solutions \cite{Kari,kkl},
\begin{equation}
Q_{\rm F}= \frac{n \ (1-(-1)^m)}{4} \, .
\end{equation}
For vanishing gravity and $m=n=1$ the Ansatz yields the 
Klinkhamer-Manton sphaleron \cite{km}.
For $n>1$ or $m>1$,
the functions $H_1,\dots,H_4$, $\Phi_1$, and $\Phi_2$
depend on $r$ and $\theta$, only. 
These axially symmetric solutions represent gravitating
multisphaleron ($m=1,n>2$), sphaleron-antisphaleron pair ($m=2,n=1$),
chain ($m>2,n=1$), and vortex ring ($m>1,n>2$) configurations
as well as mixed configurations.

With this Ansatz the full set of field equations reduces to a system 
of nine coupled partial differential equations in the independent variables 
$r$ and $\theta$. A residual U(1) gauge degree of freedom is 
fixed by the condition $r\partial_r H_1 - \partial_\theta H_2=0$ \cite{kkb}.

Regularity and finite energy require the boundary conditons 
\begin{eqnarray}
r=0: 
&  & \hspace{1.0cm}
\partial_r f(r,\theta)=
\partial_r h(r,\theta)=
\partial_r l(r,\theta)=0\, ,
\nonumber \\
&  & \hspace{1.0cm}
H_1=H_3=0\, , \ H_2=H_4=1\, ,
\nonumber \\
&  & \hspace{1.0cm}
\Phi_1=\Phi_2=0 \, , 
\nonumber \\
r\rightarrow \infty: 
&  &  \hspace{1.0cm}
f=h=l=1\, ,
\nonumber \\
&  & \hspace{1.0cm}
H_1=H_3=0\, , \ H_2=1-2m \, ,  \ 
                       1-H_4=\frac{2\sin m\theta}{\sin\theta} \, , 
\nonumber \\
&  & \hspace{1.0cm}
\Phi_1= 1 \, , \ \Phi_2=0 \, ,  
\nonumber \\
\theta = 0,\pi: 
&  & \hspace{1.0cm} 
\partial_\theta f = \partial_\theta h = \partial_\theta l =0
\nonumber \\
&  & \hspace{1.0cm}
H_1=H_3=0\, ,  \ \partial_\theta H_2=\partial_\theta H_4=0 \, ,
\nonumber \\
&  & \hspace{1.0cm}
\partial_\theta \Phi_1=0  \, , \Phi_2=0 \, ,  
\end{eqnarray}
for odd $m$, while at $r=0$
$ (\sin m\theta \Phi_1+\cos m\theta \Phi_2) = 0 $, 
$ \partial_r (\cos m\theta \Phi_1-\sin m\theta \Phi_2) = 0 $ 
is required for even $m$.

We note that the gauge potential approaches a pure gauge at infinity,
\begin{equation}
A_\mu \rightarrow \frac{i}{e} (\partial_\mu U) U^\dagger \ , \  \ \
U = e^{-i k \theta \tau_\varphi^{(n)}} \  ,
\label{Uinf} \end{equation}
when the Higgs field is in the doublet representation ($m=k$),
and likewise 
for monopole-antimonopole systems with vanishing total magnetic charge
($m=2k$), and for pure EYM solutions.

Let us now introduce the dimensionless coordinate $x$
and the dimensionless coupling constant $\alpha$ 
\begin{equation}
x = \frac{e\alpha}{\sqrt{4 \pi G}} \, r \ , \ \ \
\alpha = \sqrt{4 \pi G} v \   \ \ \
\ . \end{equation}
The limit $\alpha \to 0$ can be approached in two different
ways: 1.~$G \to 0$,
while the Higgs vacuum expectation value $v$ remains finite
(flat-space limit),
and 2.~$v \to 0$,
while Newton's constant $G$ remains finite.
These limits are then associated with different
branches of solutions.

The dimensionless mass $M$ of the solutions is
obtained from the asymptotic expansion of the metric function $f$,
\begin{equation}
M = \frac{1}{2\alpha^2} \lim_{x\to\infty} x^2 \partial_x f
=\frac{\mu}{\alpha^2}
\ . \label{mass} \end{equation}

\section{Results}

Let us first briefly recall the new static axially symmetric solutions of
Weinberg-Salam theory (in the limit of vanishing Weinberg angle),
found recently \cite{kkl}.
We here restrict the discussion to the case of vanishing Higgs mass.
The axially symmetric solutions 
are characterized by two integers, $m$ and $n$.
(We do not consider radial excitations of the solutions,
associated with a third integer.)
The electroweak sphaleron corresponds to the special case $m=n=1$,
multisphalerons are obtained for $m=1,n>1$,
while the new electroweak solutions require $m>1$.
Like the electroweak sphaleron the new solutions are unstable,
corresponding to saddle points.

For $n \le 2$, the modulus of the Higgs field
of these solutions vanishes on $m$ discrete points
on the symmetry axis, thus these flat space solutions correspond
to sphaleron-antisphaleron pairs and chains.
For $n>2$ the solutions change character,
and the modulus of the Higgs field
vanishes on one or more rings centered around the symmetry axis.
While $n=3$ and $4$ represent a transitional regime,
where (two or more) isolated nodes on the symmetry axis and rings coexist,
for larger values of $n$ these solutions possess $[m/2]$ vortex rings
(where $[m/2]$ denotes the integer part of $m/2$).
For even $m$, these solutions are thus pure electroweak vortex ring solutions,
while for odd $m$ these solutions represent electroweak 
sphaleron-vortex ring superpositions, since here 
an isolated node at the origin is retained.

Let us now consider the coupling to gravity,
while restricting to vanishing Higgs coupling constant, $\lambda=0$.
In Fig.~\ref{f-1} we exhibit the scaled mass $\mu/\alpha$ 
of the single sphaleron ($m=n=1$) and the
multisphalerons ($m=1,n=2,\dots,5$)
versus the coupling constant $\alpha$.
Whereas the mass $M$ is finite in the flat space limit $\alpha=0$,
the scaled mass $\alpha M =\mu/\alpha$ vanishes there.

In each case, a lower branch of 
gravitating sphalerons resp.~multisphalerons
emerges from the corresponding flat space solution at $\alpha=0$,
and extends up to a maximal value of the
coupling constant, $\alpha_{\rm max}$, beyond which no
such globally regular solutions exist.
At $\alpha_{\rm max}$ the lower branch
merges with a second branch, the upper branch,
which extends back to $\alpha \rightarrow 0$.
The mass $M$ diverges on the upper branch 
in the limit $\alpha \rightarrow 0$.
The scaled mass $\mu/\alpha$, in contrast, assumes a finite
limiting value.
Rescaling the solutions then shows, that
in the limit $\alpha \rightarrow 0$ in each case
a globally regular EYM solution is reached,
corresponding to the first BM solution ($n=1$)
or a generalized BM solution ($n>1$) \cite{bm,kk}.
(Note, that the $n=1$ gravitating sphaleron
was studied before \cite{greene,vg,yves}.)

\begin{figure}[h!]
\lbfig{f-1}
\begin{center}
\hspace{0.0cm}    \hspace{-0.6cm}
\includegraphics[height=.35\textheight, angle =0]{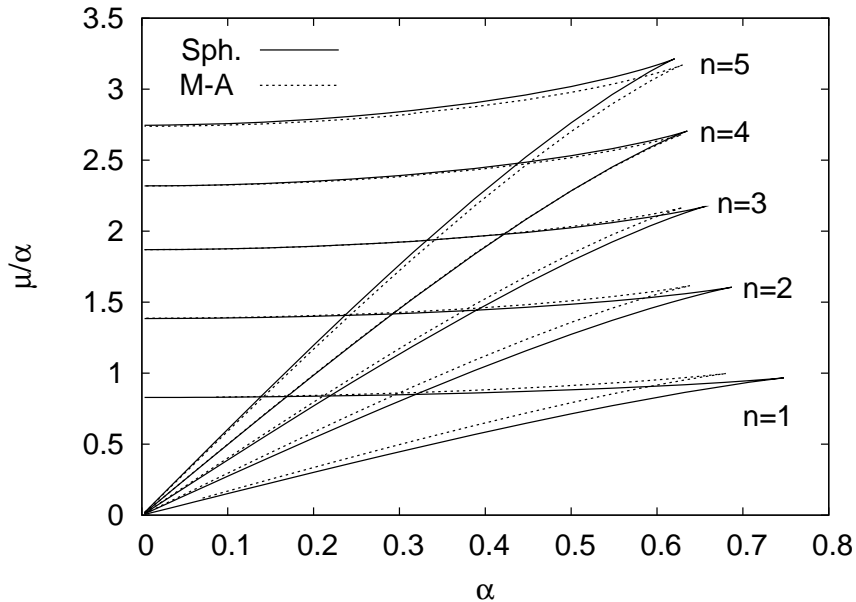}
\end{center}
\vspace{-0.5cm}
\caption{\small
Scaled mass $\mu/\alpha$ versus coupling constant $\alpha$
for the single sphaleron ($m=1,\ n=1$) 
and multisphaleron ($m=1,\ n=2,\dots,5$) solutions;
for comparison the mass of the
monopole-antimonopole pair ($m=2,\ n=1,2$) and
vortex ring ($m=2,\ n=3,5$) solutions is also shown.
}
\end{figure}

Considering the sets of solutions for different values of $n$,
we note, that with increasing $n$
the maximal value of the coupling constant $\alpha_{\rm max}$
decreases monotonically.
At the same time the maximal value of the scaled mass
increases monotonically and almost linearly.

We now compare these gravitating sphaleron and multisphaleron solutions
with gravitating monopole-antimonopole pair 
and vortex ring solutions \cite{map,kks-g}.
In these solutions the Higgs field is in the triplet representation,
thus providing only two of the gauge bosons with mass, while the
third remains massless and is therefore associated with the
electromagnetic field.
The monopole-antimonopole pair and vortex ring solutions
with $m=2$ are magnetically neutral, 
but possess magnetic dipole moments,
just like the sphaleron solutions at finite Weinberg angle 
\cite{km,kkb,hind}.

The figure shows, that
the monopole-antimonopole pair and vortex ring solutions with $m=2$
exhibit precisely the same pattern as the sphaleron 
and multisphalerons solutions with $m=1$, when their dependence on
the coupling constant $\alpha$ and the integer $n$ is considered.
Interestingly, for the case $n=4$, the agreement of the scaled mass
is excellent on both branches of solutions,
which therefore almost coincide.
Whether this is accidental, or whether there is a hidden reason
for this remarkable agreement is not clear at the moment, though.

Let us next consider larger values of $m$, i.e.,
consider the dependence of sphaleron-antisphaleron pairs
and vortex rings on the coupling constant $\alpha$.
In Fig.~\ref{f-2} we exhibit the scaled mass $\mu/\alpha$
of the gravitating sphaleron ($m=n=1$) together with the scaled mass 
of the gravitating sphaleron-antisphaleron pair ($m=2,n=1$)
and chain ($m=3,n=1$)
versus the coupling constant $\alpha$.
Again, the mass $M$ is finite in the flat space limit $\alpha=0$,
while the scaled mass $\alpha M =\mu/\alpha$ vanishes there.

\begin{figure}[h!]
\lbfig{f-2}
\begin{center}
\hspace{0.0cm}    \hspace{-0.6cm}
\includegraphics[height=.35\textheight, angle =0]{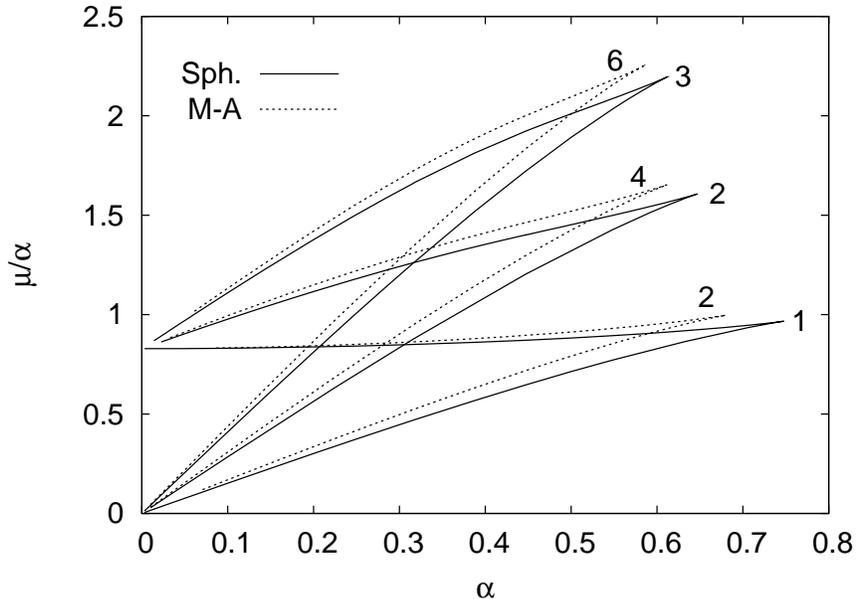}
\end{center}
\vspace{-0.5cm}
\caption{\small
Scaled mass $\mu/\alpha$ versus coupling constant $\alpha$
for single sphaleron ($m=1,\ n=1$), 
sphaleron-antisphaleron pair ($m=2,\ n=1$), and
chain ($m=3,\ n=1$) solutions;
for comparison the mass of the
monopole-antimonopole pair ($m=2,\ n=1$), and
chain ($m=4,\ n=1$),
($m=6,\ n=1$) solutions is also shown.
}
\end{figure}

Also for the sphaleron-antisphaleron pair ($m=2$)
and chain ($m=3$) a lower branch of
gravitating solutions
emerges from the corresponding flat space solution at $\alpha=0$,
and extends up to a maximal value of the
coupling constant, $\alpha_{\rm max}$,
where it merges with a second branch, the upper branch,
which extends back to $\alpha \rightarrow 0$.
In the limit $\alpha \rightarrow 0$,
the scaled mass $\mu/\alpha$ assumes the same finite
limiting value for the sphaleron-antisphaleron pair and chain
as for the single sphaleron.
Clearly, in all cases (after rescaling) the first BM solution ($n=1$)
is reached \cite{bm}.

Associating again these gravitating sphaleron-antisphaleron solutions
with the corresponding gravitating monopole-antimonopole systems,
we must resort to $m=4$ and $m=6$ monopole-antimonopole chains,
respectively.
This is clear, because we already associated a monopole-antimonopole
pair with a single sphaleron. Thus we must associate
two monopole-antimonopole pairs with the sphaleron-antisphaleron pair, 
and three with the sphaleron-antisphaleron ($m=3$) chain.
These then represent monopole-antimonopole chains
composed of $m=4$ and $m=6$ constituents, respectively,
i.e.~$m$ monopoles and antimonopoles, 
alternating on the symmetry axis.

The figure again reveals, that
the monopole-antimonopole chains with $m=4$ and $m=6$
exhibit precisely the same pattern as the sphaleron-antisphaleron
pairs ($m=2$) and chains ($m=3$), respectively, 
when their dependence on
the coupling constant $\alpha$ is considered.
In general, we conclude, that
sphaleron-antispaleron chains with $m$ constituents,
can be associated with monopole-antimonopole chains,
composed of $2m$ constituents.

Let us finally consider gravitating
sphaleron-antisphaleron vortex rings for $m=2$ and $n=4$.
In Fig.~\ref{f-3} we exhibit for these solutions
the scaled mass $\mu/\alpha$ versus the coupling constant $\alpha$,
together with the scaled mass of the gravitating multisphalerons
($m=1$, $n=4$).
For these gravitating sphaleron-antisphaleron vortex rings
we observe four instead of two branches of solutions.
As before the lower flat space branch ends and merges with the upper
branch at $\alpha_{\rm max}$. The upper branch, however,
does not connect to the generalized BM solution 
in the limit $\alpha \rightarrow 0$ \cite{kk}.
Instead it reaches an EYM solution of a different type,
which exists only when $m\ge 2$ and $n\ge 4$
\cite{ikks}.

\begin{figure}[h!]
\lbfig{f-3}
\begin{center}
\hspace{0.0cm}    \hspace{-0.6cm}
\includegraphics[height=.35\textheight, angle =0]{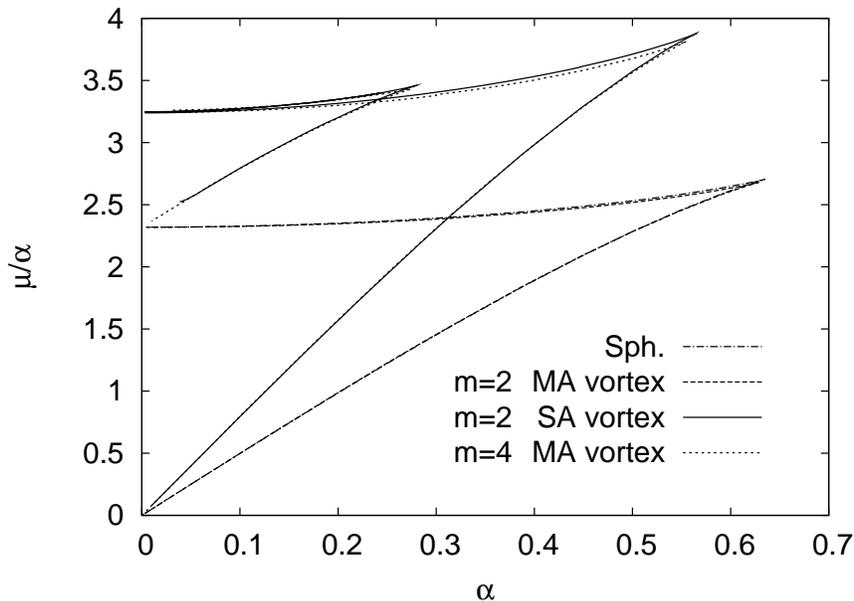}
\end{center}
\vspace{-0.5cm}
\caption{\small
Scaled mass $\mu/\alpha$ versus coupling constant $\alpha$
for single sphaleron ($m=1,\ n=4$) and
sphaleron-antisphaleron vortex ring ($m=2,\ n=4$) solutions;
for comparison the mass of
monopole-antimonopole ($m=2,\ n=4$) and
monopole-antimonopole ($m=4,\ n=4$) vortex ring solutions is also shown.
}
\end{figure}

Since EYM solutions of this type always come in pairs
for a given set of integers $m$ and $n$,
a second $n=4$ EYM solution is present,
which is slightly higher in mass \cite{ikks}.
This second EYM solution constitutes the endpoint of a second
upper branch of sphaleron-antisphaleron solutions, which merges at
a second critical value of the coupling constant $\alpha_{\rm cr}$ 
with a second lower branch,
and it is along this second lower branch,
that in the limit $\alpha \rightarrow 0$ the solutions
connect to the generalized BM solution with $n=4$.

We now compare these gravitating $m=2$
sphaleron-antisphaleron vortex ring solutions again
with the corresponding gravitating $m=4$ monopole-antimonopole 
vortex ring solutions.
As anticipated, the figure reveals, that
the monopole-antimonopole vortex rings with $m=4$
exhibit precisely the same pattern as the sphaleron-antisphaleron
vortex rings with $m=2$, when their dependence on
the coupling constant $\alpha$ is considered.
Moreover, the masses show again an intriguing
quantitative agreement for these $n=4$ solutions,
which remains to be understood.

The sphaleron-antisphaleron chain and vortex ring configurations
presented above were all obtained for vanishing Higgs mass,
As a function of the Higgs mass these solutions also form branches
\cite{kkl}.
At critical values of the Higgs mass bifurcations arise, 
where new branches of solutions appear. 
These then give rise to a plethora of further gravitating solutions
to be discussed elsewhere.

\section{Conclusions}

We have investigated gravitating sphalerons, multisphalerons
and sphaleron-antisphaleron systems, which are static and
axially symmetric, and characterized by two integers, $m$ and $n$.
Single sphalerons are obtained for $m=n=1$,
multisphalerons for $m=1$ and $n>1$,
and sphaleron-antisphaleron systems for $m>1$.
Like the electroweak sphaleron these new solutions are unstable,
corresponding to saddle points.

In the presence of gravity, 
from each of these flat space solutions,
a branch of gravitating solutions emerges
and evolves smoothly with increasing 
gravitational coupling constant $\alpha$
up to a maximal value $\alpha_{\rm max}$.
There it merges with a second branch,
higher in energy,
which extends backwards to $\alpha=0$.
In the limit, the Higgs vacuum expectation value vanishes,
and the limiting solutions correspond to
pure EYM solutions (after rescaling).

For larger values of the Higgs mass,
the flat space solutions are no longer uniquely specified by the
integers $m$ and $n$. Instead bifurcations appear, giving rise to
further branches and types of configurations.
As for the monopole-antimonopole systems \cite{kks-g},
we therefore expect a plethora of gravitating solutions
at large scalar coupling.
Furthermore, for very large values of the Higgs mass also
bisphalerons or `deformed' sphalerons are present \cite{bi,yves},
which do not exhibit parity reflection symmetry.

Comparing these gravitating sphaleron, multisphaleron 
and sphaleron-antisphaleron solutions, based on a doublet Higgs field,
with the monopole-antimonopole solutions, obtained with
a triplet Higgs field,
we find precisely the same pattern of branches of solutions,
when we compare sphalerons and sphaleron-antisphaleron systems
characterized by $m$ and $n$,
with monopole-antimonopole systems 
characterized by $2m$ and $n$.
Interestingly, in the case $n=4$, the scaled mass
of both types of solutions even almost coincides.

Monopole-antimonopole systems can rotate,
when they carry no magnetic charge \cite{radu}.
It therefore appears interesting to consider also rotating
sphaleron-antisphaleron systems.
Moreover, monopole-antimonopole systems can be endowed with
a black hole at their center, as shown explicitly already for the
monopole-antimonopole pair \cite{mapbh}.
Sphaleron-antisphaleron systems with black holes are thus expected
to exist as well.

{\sl Acknowledgement}

R.I. gratefully acknowledges support by the Volkswagen Foundation,
and B.K. support by the DFG.

\end{document}